\documentclass[aps,prl,nofootinbib,twocolumn,preprintnumbers,floatfix,superscriptaddress, bibliography]{revtex4-1}
\usepackage[linkcolor=red, anchorcolor=blue, citecolor=blue, colorlinks=true, urlcolor=blue]{hyperref}
\usepackage[figuresright]{rotating}
\usepackage{amsmath}
\usepackage{graphicx}
\usepackage{ulem}
\usepackage{bm}
\usepackage{xspace}
\usepackage{amssymb}
\usepackage{gensymb}
\usepackage{multirow}
\usepackage{threeparttable}
\usepackage{amsthm}
\usepackage{mathrsfs}
\usepackage{indentfirst}
\usepackage{epstopdf}
\usepackage{breakurl}
\usepackage{cancel}
\usepackage{slashed}

\begin{document}

\title{Hierarchical Axiverse}

\author{Hai-Jun Li} 
\email{lihaijun@ynu.edu.cn}
\affiliation{Department of Physics, Yunnan University, Kunming 650091, China}

\preprint{YNU-HEP-26-034}

\date{\today}

\begin{abstract}

String compactifications generically produce ${\cal O}(100)$ light axion-like particles, yet low-energy mixing has not previously been exploited as a structuring constraint on their spectrum.
We show that the requirement of well-defined QCD-induced mass mixing forces the axion masses into a hierarchically spaced ordering, while stochastic mixing organizes the decay constants into a hierarchically split distribution --- two populations separated by an essentially empty window.
Both patterns are insensitive to ultraviolet input and satisfy the relevant Swampland constraints within standard large-volume compactifications.
We term this emergent pattern the hierarchical axiverse.


\end{abstract}
\maketitle


\medskip\noindent{\bf Introduction.}---%
String theory compactifications generically yield a plenitude of axion fields arising from the dimensional reduction of higher-form gauge potentials on non-trivial cycles of the internal manifold~\cite{Witten:1984dg, Choi:2003wr, Svrcek:2006yi}.
The resulting low-energy spectrum --- the string axiverse --- contains ${\cal O}(100)$ axion-like particles (ALPs) whose masses can span many orders of magnitude, from the Hubble scale $H_0\sim 10^{-33}\,{\rm eV}$ up to values well above the QCD scale $\Lambda_{\rm QCD}\sim 10^8\, {\rm eV}$~\cite{Arvanitaki:2009fg, Cicoli:2012sz, Demirtas:2018akl, Reig:2021ipa, Demirtas:2021gsq, Leedom:2025mlr, Fallon:2025lvn}.
This rich particle content has far-reaching consequences for particle physics and cosmology: the QCD axion provides an elegant solution to the strong CP problem~\cite{Peccei:1977hh, Peccei:1977ur, Weinberg:1977ma, Wilczek:1977pj}, both the QCD axion and ultra-light ALPs are viable dark matter (DM) candidates~\cite{Preskill:1982cy, Abbott:1982af, Dine:1982ah}, and both are targets of ongoing haloscope, helioscope, and laboratory searches~\cite{Marsh:2015xka, OHare:2024nmr}.
A central question for the axiverse is whether, in addition to being fixed by ultraviolet (UV) input such as compactification geometry and instanton dynamics, the mass and coupling patterns of this large spectrum can also emerge from the low-energy mixing structure itself.
 
In a multi-axion system, the physical states are not the individual lattice fields but linear combinations thereof, related by both kinetic mixing (from the off-diagonal Kähler metric) and mass mixing (from the temperature-dependent QCD axion potential that turns on as the Universe cools).
Previous studies have employed axion mixing primarily as a phenomenological tool: most directly, it redistributes the DM abundance and isocurvature fluctuations among the QCD axion and ALPs~\cite{Daido:2015cba, Li:2023uvt, Cyncynates:2023esj}, and more broadly modifies other cosmological and astrophysical observables~\cite{Li:2025cep, Daido:2015bva, Ho:2018qur, Cyncynates:2022wlq, Kitajima:2023cek, Lee:2024xjb, Murai:2024nsp, Li:2025imm, Murai:2025wbg}.
The axion mass spectrum itself, however, is conventionally treated as an input fixed by non-perturbative instanton dynamics~\cite{Baryakhtar:2026oun}; whether the mixing structure alone can organize the spectrum has received comparatively little attention.
Moreover, the recently proposed stochastic mixing mechanism~\cite{Li:2026rga} --- in which effective mixing can occur regardless of the relative magnitudes of the axion decay constants --- opens up parameter space beyond the deterministic maximal mixing limit considered in earlier work.

In this Letter, we show that axion mixing generically produces hierarchical structures in both the mass spectrum and the decay constant distribution, and that these patterns are intrinsic consequences of the mixing dynamics rather than inputs from the UV completion.
The requirement of well-defined, non-degenerate mixing forces the ALP masses into a hierarchically spaced ordering $m_{A_i}/m_{A_{i+1}}\lesssim 1/2$, while stochastic mixing drives the decay constant ratio $R_f\equiv f_{A_i}/f_{a_0}$ into two separated populations with a characteristic gap $0.1\lesssim R_f\lesssim 10$ left essentially empty.
Unlike hierarchies attributed to non-perturbative instanton dynamics~\cite{Baryakhtar:2026oun}, both patterns emerge purely from the mixing structure.
We further verify that the resulting spectrum is compatible with standard Swampland~\cite{Vafa:2005ui} constraints --- in particular, the axion Weak Gravity Conjecture~\cite{Arkani-Hamed:2006emk, Harlow:2022ich} and the axion Festina Lente bound~\cite{Montero:2019ekk, Montero:2021otb, Guidetti:2022xct} --- without requiring additional model-building ingredients.

\medskip\noindent{\bf Axiverse kinetic mixing.}---%
In the string axiverse, the effective axion theory is initially formulated in a lattice basis $\theta^i$ derived from compactification geometry. The relevant Lagrangian is
\begin{eqnarray}
\mathcal{L}\supset\dfrac{1}{2}K_{ij}\partial_\mu\theta^i \partial^\mu\theta^j-\sum_I\Lambda_I^4\left[1-\cos\left(\sum_i {\mathcal{Q}^I}_i \theta^i\right)\right] ,
\end{eqnarray} 
where $K_{ij}$ is the Kähler metric, $\Lambda_I$ the instanton scale, ${\mathcal{Q}^I}_i$ the charge matrix, and the sum runs over all instanton sectors (the constant phase is set to zero).
To obtain canonical kinetic terms $\frac{1}{2}\partial_\mu\phi^i\,\partial^\mu\phi^i$, we redefine $\phi^i={\rm diag}(f_K)^i{}_k\,{U^k}_j\,\theta^j$, where $f_K=\sqrt{{\rm eig}(K_{ij})}$ and ${U^k}_j$ diagonalizes $K_{ij}$.
In this basis the Hessian reads $\mathcal{H}_{ij}^\phi={F^m}_p\,{U^p}_i\,{F^n}_q\,{U^q}_j\,\mathcal{H}_{mn}^\theta$, with ${F^i}_j={\rm diag}(1/f_K)^i{}_j$. Diagonalizing $\mathcal{H}_{ij}^\phi$ with eigenvector matrix ${T^i}_j$ yields the mass eigenstates $\varphi^i={T^i}_j\,\phi^j$ and physical masses $m_i=\sqrt{{\rm eig}(\mathcal{H}^\phi)}$.
The quartic coupling tensor transforms to the mass eigenbasis as $\lambda_{ijkl}^\varphi={T^a}_i\,{T^b}_j\,{T^c}_k\,{T^d}_l\,\lambda_{abcd}^\phi$, from which the perturbative decay constants follow as $f_{i,\rm pert}^2=m_i^2/|\lambda_{iiii}^\varphi|$ (the Taylor expansion factor is absorbed into the definition).

With the mass spectrum and perturbative decay constants in hand, we turn to the interactions of the physical axions. 
Since the mass eigenstates $\varphi^i$ are linear superpositions of the lattice fields, the coupling of each physical axion to an external sector is the corresponding linear combination of the lattice couplings, weighted by the rotation matrix elements ${T^i}_j$.
The treatment of axion mass mixing, however, calls for a physically distinct framework, to which we now turn.

\medskip\noindent{\bf Axiverse mass mixing.}---%
Mass mixing is driven by the QCD-induced temperature-dependent axion mass, while the ALP masses are assumed to be temperature-independent.
Specifically, the QCD axion effective potential is given by 
\begin{align}
V_{\rm QCD}=m_{a_0}^2(T) f_{a_0}^2\left[1-\cos\left(\theta_{\rm QCD} \right)\right]\, ,
\end{align} 
where $m_{a_0}$ and $f_{a_0}$ denote the axion mass and decay constant, respectively.
The temperature dependence of the QCD axion mass follows
\begin{align}
m_{a_0}(T)=m_{a_0,0}
\begin{cases}
1\, , &T\leq T_{\rm QCD} \\ 
\left(T/T_{\rm QCD}\right)^{-b}\, , &T>T_{\rm QCD}
\end{cases}
\end{align}
where $T_{\rm QCD}\simeq150\, \rm MeV$, $m_{a_0,0}$ is the zero-temperature QCD axion mass, and $b\simeq 4.08$ as determined by lattice QCD simulations.
Incorporating this temperature dependence into the multi-axion system, the relevant Lagrangian reads
\begin{eqnarray}
\mathcal{L}\supset\dfrac{1}{2}\sum_{i=0}^N f_i^2\left(\partial\theta^i\right)^2-\sum_{i=0}^N\Lambda_i^4\left[1-\cos\left(\sum_{j=0}^N {n^i}_j \theta^j\right)\right]  ,
\end{eqnarray} 
where $f_i$ is the axion decay constant and ${n^i}_j$ is the domain wall number matrix.
Unlike the lattice basis $\theta^i$ used in the kinetic mixing analysis, here $\theta^i$ already denotes the physical fields before QCD-induced mixing is turned on.
The index $i=0,1,\ldots,N$ labels the $N+1$ axion fields, where $a_0$ is the QCD axion and $A_i$ ($i=1,\ldots,N$) denote the ALPs, with masses $m_{a_0}$ and $m_{A_i}$, respectively.
Furthermore, we label the ALPs such that $m_{A_i}<m_{A_{i+1}}<m_{a_0,0}$.
The parameters are defined as $f_i = f_{a_0}$ and $\Lambda_i = \sqrt{m_{a_0} f_{a_0}}$ for $i=0$, while $f_i = f_{A_i}$ and $\Lambda_i = \sqrt{m_{A_i} f_{A_i}}$ for $i \geq 1$.
As a technical prerequisite, each ALP decay constant should lie outside the comparable regime $f_{A_i}\sim f_{a_0}$; this prevents eigenvalue splitting in the intermediate-temperature regime and, although not strictly necessary for the $N=1$ case, is essential for $N\geq 2$~\cite{Li:2025cep}.
The domain wall number matrix ${n^i}_j$ is given by
\begin{align}
{n^i}_j = \delta_{ij}+d_j \delta_{i0}(1-\delta_{j0})+(1-d_i) \delta_{j0}(1-\delta_{i0})\, ,
\end{align}
where $\delta_{ij}$ is the Kronecker delta, $d_i=0$ if $f_{A_i}\ll f_{a_0}$, and $d_i=1$ if $f_{A_i}\gg f_{a_0}$.
The label $d_i$ merely records the relative magnitude of each ALP decay constant with respect to $f_{a_0}$, with no global alignment imposed among them.
Note that we adopt a stochastic mixing mechanism, which allows effective mixing to occur across all $N$ ALPs regardless of the relative magnitudes of their decay constants~\cite{Li:2026rga}.
This cannot be realized in the previous maximal mixing mechanism, where the effective number of mixings would be reduced. 
Specifically, in the maximal mixing limit, one obtains two distinct scenarios with domain wall number matrices 
\begin{align}
{n^i}_j&=\delta_{j0}+\delta_{ij}(1-\delta_{i0})\, ,\\
{n^i}_j&=\delta_{i0}+\delta_{ij}(1-\delta_{j0})\, ,
\end{align} 
which illustrates maximal mixing as a deterministic limit of stochastic mixing.

While axion mass mixing leads to rich phenomenological consequences~\cite{Li:2025cep}, our focus is on its implications for the axiverse properties themselves.

\begin{figure*}[t]
\centering
\includegraphics[width=0.48\textwidth]{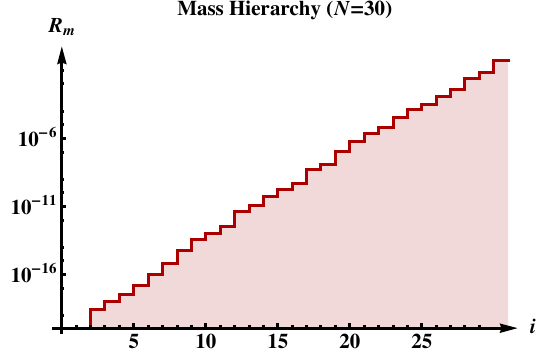}\quad\includegraphics[width=0.48\textwidth]{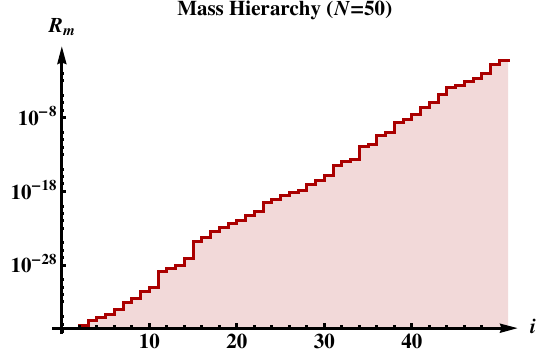}  
\caption{Hierarchical axion mass distribution in the axiverse with $N=30$ (left) and $N=50$ (right).
The horizontal axis labels the ALP index $i$.
The vertical axis shows the mass ratio $R_m\equiv m_{A_i}/m_{a_0,0}$, where $m_{A_i}$ is the mass of the $i$-th ALP and $m_{a_0,0}$ is the zero-temperature mass of the QCD axion.}
\label{fig_mass_distribution}
\end{figure*}

\medskip\noindent{\bf Hierarchical axiverse.}---%
We now show that axion mixing generates hierarchical structures in the axiverse, primarily in the mass and decay constant distributions.
We define the hierarchical axiverse as the emergent, UV-insensitive pattern in which (i)~the axion masses are hierarchically spaced by the requirement of well-defined mixing, and (ii)~the decay constants are split into two populations separated by an empty window.

We begin with the axion mass distribution.
Two physically distinct requirements constrain the ALP spectrum.
First, no two ALPs may be exactly degenerate in mass.
If there exist $i\neq j$ such that $m_{A_i}=m_{A_j}$, the two fields combine into a single effective degree of freedom and only one participates in the QCD-induced mixing; the mixing matrix would then be ill-defined.
Non-degeneracy, $m_{A_i}\neq m_{A_j},\, \forall\, i\neq j$, is therefore a prerequisite for well-defined mass mixing.
Second, in a multi-axion system the temperature-dependent QCD axion mass sweeps through the ALP spectrum as the Universe cools. 
From the temperature-dependent mass matrix, we obtain the heavy and light eigenmasses $m_{h_i,l_i}$ and solve $\left(m_{h_i}-m_{l_i}\right)|_{T\to 0}=\left(m_{h_i}-m_{l_i}\right)|_{T\to +\infty}$ to derive the mass factor $1/2$.
For each mass crossing to be individually resolvable --- $\rm i.e.$, for the crossings not to overlap --- adjacent masses should satisfy $m_{A_i}/m_{A_{i+1}}\lesssim 1/2$ and $m_{A_i}/m_{a_0,0} \lesssim \left(1/2\right)^{N-i+1}$, where this factor $1/2$ serves as the reference value for moderate decay constant ratios $f_{A_i}/f_{a_0}={\cal O}(0.1\text{--}1)$; this bound tightens (loosens) for larger (smaller) ratios.
Under the non-degeneracy condition, we may label the ALPs in order of increasing mass, $m_{A_1}<m_{A_2}<\cdots<m_{A_N}<m_{a_0,0}$; this ordering is a convention, not an additional assumption.
The physical content resides entirely in the two requirements above: they force the spectrum to be hierarchically spaced, with each successive ALP approximately a factor of two heavier than the preceding one.

This mass distribution is illustrated schematically in Fig.~\ref{fig_mass_distribution}.
In the left panel, we show the $R_m$ distribution for $N=30$, where $R_m\equiv m_{A_i}/m_{a_0,0}$. 
As $i$ increases, $R_m$ increases monotonically, reflecting the imposed mass ordering $m_{A_i} < m_{A_{i+1}}$.
In the right panel, we present the case for $N=50$, where a similar distribution is observed.
As $N$ increases, the lower end of the $R_m$ distribution extends to smaller values, since the lightest ALP satisfies $m_{A_1}/m_{a_0,0} \lesssim \left(1/2\right)^N$; the spectrum thus spans a progressively wider range.
A clear hierarchy is evident in the $R_m$ distribution, reflected in the spacing between adjacent ALPs in the mass spectrum. 
For visual clarity, a lower cutoff $m_{A_i}/m_{A_{i+1}}\gtrsim10^{-1}$ is applied to the figure; it does not reflect a physical constraint. 

Within a similar string axiverse framework, a hierarchical mass spectrum has also been noted in Ref.~\cite{Baryakhtar:2026oun}, where it is attributed to hierarchical instanton effects with statistically distributed axion couplings. 
In our setup, however, the hierarchy is not inherited from the underlying non-perturbative dynamics but is a necessary consequence of the mixing structure: it is the requirement of well-defined mass mixing that organizes the hierarchical spectrum.
The two origins are physically distinct: in what follows, the hierarchy should be understood as a consequence of mixing rather than as an input assumption of the axiverse construction.

We now turn to the distribution of axion decay constants.
In the stochastic mixing mechanism, no global alignment of the $d_i$ labels is imposed: each ALP independently satisfies either $f_{A_i}\ll f_{a_0}$ or $f_{A_i}\gg f_{a_0}$, as encoded in the domain wall number matrix above.
This contrasts with earlier two-axion models~\cite{Daido:2015cba,Li:2023uvt,Cyncynates:2023esj}, in which a single hierarchy is imposed by construction, and with the multi-axion extension~\cite{Li:2025cep}, where all ALPs share the same sign of $d_i$.
In those constructions no ordering among the ALPs themselves is generated; here it emerges from stochastic mixing.
Unlike the ALP masses, the decay constants $f_{A_i}$ can be degenerate among themselves; nevertheless, a hierarchy between $f_{a_0}$ and each $f_{A_i}$ generically emerges from the mass mixing.

\begin{figure*}[t]
\centering
\includegraphics[width=0.48\textwidth]{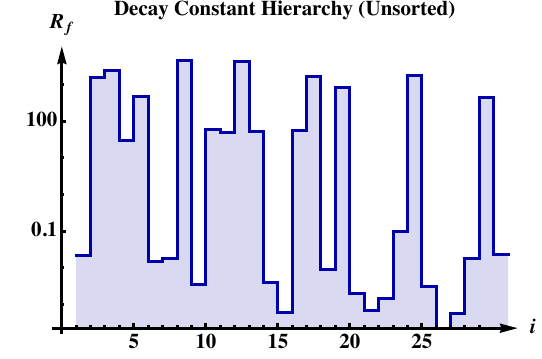}\quad\includegraphics[width=0.48\textwidth]{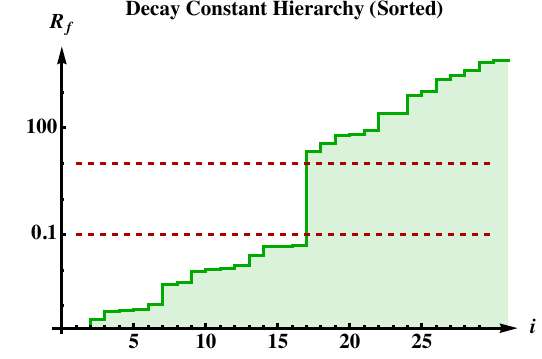}  
\caption{Hierarchical axion decay constant distribution in the axiverse with $N=30$, comparing the unsorted (left) and sorted (right) cases.
The vertical axis shows the ratio $R_f\equiv f_{A_i}/f_{a_0}$, where $f_{A_i}$ is the decay constant of the $i$-th ALP and $f_{a_0}$ is that of the QCD axion.
The horizontal axis labels the ALP index $i$; in the right panel the ALPs are reordered by increasing $R_f$.
The two red dashed lines mark the interval $0.1\lesssim R_f\lesssim 10$.}
\label{fig_decay_constant_distribution}
\end{figure*} 

\begin{table}[t]
\caption{Comparison of the hierarchical structures in axion masses and decay constants.}
\begin{ruledtabular}
\begin{tabular}{lcc}
Property  & Mass & Decay constant   \\
\hline
Origin           & Mixing dynamics     & Mixing dynamics   \\
Manifestation    & Full spectrum       & Gap at $0.1\lesssim R_f\lesssim 10$ \\
Degeneracy\tablenote{Entries indicate whether degeneracy is allowed for QCD axion--ALP / ALP--ALP cases.} & Absent / Absent          & Absent / Present  \\
UV sensitivity   & Insensitive         & Insensitive       \\
\end{tabular}
\end{ruledtabular}
\label{tab_1}
\end{table}

The resulting decay constant distribution for $N=30$ is illustrated in Fig.~\ref{fig_decay_constant_distribution}. 
The left panel shows the unsorted ratio $R_f\equiv f_{A_i}/f_{a_0}$, while the right panel displays the sorted distribution, in which the hierarchy among the ALPs becomes more pronounced.
A striking feature is the absence of data in the interval $0.1\lesssim R_f\lesssim 10$, bounded by the red dashed lines: stochastic mixing generically drives $R_f$ either well below or well above unity, leaving the intermediate range essentially empty.
The decay constant hierarchy therefore manifests itself across this gap --- between ALPs with $R_f\ll 1$ and those with $R_f\gg 1$ --- rather than among neighbors on the same side, in contrast to the mass hierarchy shown in Fig.~\ref{fig_mass_distribution}, which extends across the full spectrum.
Crucially, the decay constant hierarchy arises not only between the QCD axion and the ALPs but also among the ALPs themselves, driven by stochastic mixing rather than imposed by construction.

Note that the hierarchical properties of axion masses and decay constants here are addressed within the context of axion mixing.
We summarize the comparison in Table~\ref{tab_1}.
Both hierarchies originate from the same mixing dynamics, yet they differ in two key aspects.
First, the mass hierarchy spans the full spectrum, whereas the decay constant hierarchy resides in the contrast across the gap at $0.1\lesssim R_f\lesssim 10$.
Second, exact mass degeneracy between any pair of axions is forbidden, as it would spoil the well-defined mass mixing.
Decay constant degeneracy between the QCD axion and ALPs is likewise excluded, as it nullifies the effective mixing and lies outside the present framework.  
The only allowed degeneracy is that of decay constants among ALPs themselves.
Moreover, neither hierarchy depends on UV details. 
 
\medskip\noindent{\bf Swampland consistency.}---%
We further verify compatibility with the Swampland bounds tied to axion decay constants; these serve as consistency checks and do not generate the hierarchies above.
As for the mass hierarchy, it is automatically safe because $m_{A_i}<m_{a_0,0}$, and no bound restricts the mass ratios directly.

The axion Weak Gravity Conjecture requires, up to ${\cal O}(1)$ factors, an instanton of action $S_i$ for each axion such that $f_iS_i\lesssim M_{\rm Pl}$~\cite{Arkani-Hamed:2006emk, Harlow:2022ich}, where $M_{\rm Pl}$ is the reduced Planck mass.
For the $R_f\ll1$ population this condition is immediate, since $f_{A_i}\ll f_{a_0}\lesssim M_{\rm Pl}$ and an ${\cal O}(1)$ instanton action already suffices.
For the $R_f\gg1$ population, the same condition is realized in
controlled large-volume compactifications, where a simple estimate
gives $f_i\sim M_{\rm Pl}/{\cal V}^{2/3}$ and
$S_i\sim{\cal V}^{2/3}$, with ${\cal V}$ the compactification
volume in string units.  Then
\begin{align}
f_iS_i\sim M_{\rm Pl}\lesssim M_{\rm Pl}\, ,
\end{align}
saturating the bound, so the upper population is not in tension
with the Weak Gravity Conjecture.
Since the conjecture constrains $f_iS_i$ rather than $R_f$ itself,
it neither explains nor forbids the mixing-induced gap.

The axion Festina Lente bound gives the complementary condition $f_iS_i\gtrsim\sqrt{M_{\rm Pl}H}\sim\rho_\Lambda^{1/4}$~\cite{Montero:2019ekk, Montero:2021otb, Guidetti:2022xct}, where $H$ is the Hubble scale of the quasi-de Sitter background and $\rho_\Lambda\simeq3M_{\rm Pl}^2H^2$.
Evaluated at the present Hubble scale $H_0$, the lower edge is ${\cal O}(10^{-3})\,{\rm eV}$, far below the typical value of $f_iS_i$ in controlled compactifications.
Combining the two bounds yields
\begin{align}
\sqrt{M_{\rm Pl}H_0}\lesssim f_iS_i\lesssim M_{\rm Pl}\, .
\end{align}
The large-volume axiverse naturally realizes this window: $f_iS_i\sim M_{\rm Pl}$ saturates the upper bound but remains many orders of magnitude above the lower bound for controlled compactification volumes.

Together, these checks confirm that the hierarchical axiverse is compatible with the relevant Swampland bounds within the standard large-volume setup.

\medskip\noindent{\bf Conclusion.}---%
We have shown that the hierarchical ordering of the string axiverse --- in both masses and decay constants --- is not merely an assumption imported from compactification geometry or instanton dynamics, but a consistency condition of the mixing itself.
Once well-defined mass mixing is required, the axion mass ordering and the decay constant gap follow.

This reframing has immediate phenomenological consequences.
The hierarchically spaced spectrum modifies the standard DM abundance calculation: the relic density is redistributed among non-degenerate states whose spacings are fixed by the mixing conditions rather than chosen freely~\cite{Li:2025uwq}.
Isocurvature constraints, which scale with the number of light degrees of freedom during inflation, become more restrictive as $N$ grows and the lightest ALP mass is driven to lower values.
A dedicated study of these effects will be presented elsewhere.

Embedding the construction in high-scale inflation would strengthen the axion Festina Lente lower bound; this extension deserves separate treatment.
Finally, the empty window in the decay constant ratio constitutes a falsifiable prediction: observation of an ALP whose decay constant ratio falls within this range would challenge the purely mixing-origin scenario and point to additional UV contributions.
 
\medskip\noindent{\bf Acknowledgments.}---%
This work was supported by Yunnan University and the Key Laboratory of Astroparticle Physics of Yunnan Province.
 
\bibliography{references}
\end{document}